\documentclass[article]{new-aiaa}
\usepackage[utf8]{inputenc}
\usepackage{textcomp}
\usepackage{xcolor}
\usepackage{graphicx}
\usepackage{amsmath}
\usepackage[version=4]{mhchem}
\usepackage{siunitx}
\usepackage{longtable,tabularx}
\usepackage{appendix}
\usepackage{multirow}
\usepackage{float}

\usepackage{sectsty}
\usepackage{indentfirst}
\renewcommand\thesection{\arabic{section}}
\renewcommand\thesubsection{\thesection.\arabic{subsection}}
\usepackage{setspace}
\sectionfont{\fontsize{14}{12}\selectfont}
\subsectionfont{\fontsize{13}{12}\selectfont}
\subsubsectionfont{\fontsize{12}{12}\selectfont}

\title{Attention-based Multi-fidelity Machine Learning Model for Computational Fractional Flow Reserve Assessment}
\author{\fontsize{10pt}{14pt}\selectfont Haizhou Yang\textsuperscript{1}, C. Alberto Figueroa\textsuperscript{2,3}\footnote{Professor, University of Michigan, figueroc@med.umich.edu} and Krishna Garikipati\textsuperscript{1,4,5}\footnote{Professor, University of Michigan, krishna@umich.edu}}
\affil{\fontsize{10pt}{14pt}\selectfont \textsuperscript{1}Department of Mechanical Engineering, University of Michigan, Ann Arbor, MI 48109, USA}
\affil{\fontsize{10pt}{14pt}\selectfont \textsuperscript{2}Department of Biomedical Engineering, University of Michigan, Ann Arbor, MI 48109, USA}
\affil{\fontsize{10pt}{14pt}\selectfont \textsuperscript{3}Department of Surgery, University of Michigan, Ann Arbor, MI 48109, USA}
\affil{\fontsize{10pt}{14pt}\selectfont \textsuperscript{4}Department of Mathematics, University of Michigan, Ann Arbor, MI 48109, USA}
\affil{\fontsize{10pt}{14pt}\selectfont \textsuperscript{5}Michigan Institute for Computational Discovery \&  Engineering, University of Michigan, Ann Arbor, MI 48109, USA}

\begin{document}

\maketitle

\begin{abstract}
Coronary Artery Disease (CAD) is one of the most common forms of heart disease, which is caused by a buildup of atherosclerotic plaque (known as stenosis) in the coronary arteries, leading to insufficient supplement of blood, oxygen, and nutrients to the heart. Fractional Flow Reserve (FFR), measuring the pressure ratio between the aorta and distal coronary artery, is an invasive physiologic gold standard for assessing the severity of coronary artery stenosis. Despite its benefits, invasive FFR assessment is still underutilized due to its high cost, time-consuming, experimental variability, and increased risk to patients. In this study, an attention-based multi-fidelity machine learning model (AttMulFid) is proposed for computationally efficient and accurate FFR assessment with uncertainty measurement. Within AttMulFid, an autoencoder is utilized to intelligently select geometric features from coronary arteries, with additional attention on the key area. Results show that the geometric features are able to represent the entirety of the geometric information and intelligently allocate attention based on crucial properties of geometry. Furthermore, the AttMulFid is a feasible approach for non-invasive, rapid, and accurate FFR assessment (with 0.002s/simulation).
\end{abstract}

\section{Introduction}

Coronary artery disease (CAD) is one of the most common forms of heart disease characterized by the narrowing and blockage (known as stenosis) of the coronary arteries, which supply oxygen-rich blood to the heart muscle. It is associated with a wide range of clinical manifestations, including stable angina, acute coronary syndrome, myocardial infarction, heart failure, and sudden cardiac death \cite{malakar2019review,hajar2017risk}. These conditions can lead to severe disability, reduced quality of life, and premature death. CAD is the leading cause of death worldwide, accounting for millions of deaths annually \cite{alizadehsani2019machine}. Furthermore, CAD also places a considerable economic burden on healthcare systems due to the costs associated with hospitalization, diagnostic tests, medications, interventions, and long-term management of complications \cite{kumar2022cost}. In summary, CAD affects individuals across all age groups and socioeconomic backgrounds, emphasizing the need for accurate diagnosis of CAD for appropriate management and risk stratification. 

Diagnostic modalities range from non-invasive tests such as stress testing, coronary computed tomography angiography, and cardiac magnetic resonance imaging, to invasive procedures such as coronary angiography \cite{abdar2019new,tamis2019contemporary}. The non-invasive diagnostic methods offer several advantages, such as being safe, readily available, and less costly compared to invasive procedures. However, they might yield low accuracy of the results and limited spatial resolution. On the other hand, invasive procedures provide valuable information about the extent, severity, and complexity of CAD, serving as the golden standard for assessing the anatomical severity of CAD and guiding revascularization decisions. However, they are always associated with potential risks and no functional information is provided. In order to acquire functional information, fractional flow reserve (FFR) is employed to provide a physiological assessment of stenosis severity and functional significance \cite{de2008fractional}. It is defined as the pressure ratio across a stenosis ($\mathrm{FFR} = P_d/P_a$, where $P_d$ is distal coronary artery pressure, and $P_a$ is aortic pressure), measured by a catheter equipped with a pressure sensor inserted into the coronary artery. FFR-guided revascularization has been shown to improve patient outcomes, reduce the need for unnecessary interventions, and optimize resource utilization \cite{van2020usefulness}. Nevertheless, the procedure may pose challenges in terms of time-consuming, costly, experimental variability, and increased potential risks to patients \cite{fearon2019accuracy,tu2020fractional}. Therefore, non-invasive, computational approaches for measuring FFR are highly desired.

Numerous computational methods have been developed to effectively and precisely assess FFR without the need for invasive procedures. Zarins et al. and Min et al. computed FFR from patient-specific computed tomographic (CT) angiography data using computational fluid dynamics (CFD) model based on 3D navier-stokes (NS) equations \cite{zarins2013computed,min2015noninvasive}. Despite its high accuracy, high-fidelity Computational Fluid Dynamics (CFD) suffers from long computational time. Coenen et al. presented an on-site algorithm for the computation of FFR based on coronary CT scans. This algorithm employs a reduced-order model to accelerate the computation process \cite{coenen2015fractional}. Similarly, CathWorks developed a rapid computational method for FFR assessment based on angiograms utilizing a lumped parameter model \cite{pellicano2017validation}, which demonstrated promising results in clinical studies \cite{lavi2020calculating}. However, it should be noted that the computation models employed in both approaches are low-fidelity models. The simplified nature of the low-fidelity model sacrifices some precision in capturing the intricate details of the hemodynamic behavior, potentially leading to less accurate FFR assessments. As a result, the FFR results obtained may be less accurate compared to more sophisticated and high-fidelity models. 

In order to achieve both computational speed and accuracy, machine learning-based approaches are investigated for computational FFR assessment \cite{tesche2020machine}. Itu et al. proposed a machine-learning model for FFR prediction which is trained using synthetically generated coronary anatomies and a reduced-order CFD model \cite{itu2016machine}. Fossan et al. combined a physics-based reduced-order model with a fully connected neural network to estimate the pressure losses across stenotic and healthy coronary segments \cite{fossan2021machine}. Tesche et al. also demonstrated comparable performance of computational FFR assessment between machine learning-based and CFD-based approaches. Both methods exhibited superior accuracy compared to coronary CT angiography and quantitative coronary angiography in detecting stenosis \cite{tesche2018coronary}. However, the input features of these machine learning models are selected manually or lack sufficient intelligence, especially for the geometric features, making it challenging to determine if all the significant geometric features affecting pressure losses have been included. Additionally, both of them utilized fully connected neural networks to estimate FFR, without leveraging the potential benefits of utilizing feature maps that could be extracted from FFR/pressure/velocity profiles along the vessel.

In this study, an attention-based multi-fidelity machine learning model (AttMulFid) is proposed to computationally assess FFR of coronary vessels. In contrast to existing machine learning approaches, the proposed method presents several novelties: (1) The proposed approach utilizes an autoencoder to intelligently select geometric features, a low-rank representation, from coronary arteries, with additional attention on the key area, such as stenosis. (2) Gradient-based attention is computed to generate visual explanations of the feature extraction process. This attention mechanism specifically highlights the switching of attention toward key areas. (3) Uncertainty quantification is performed to provide more insight into the FFR prediction, i.e. the confidence interval. The uncertainty bound is beneficial to guide clinical decisions, such as revascularization. (4) This study represents an initial effort to integrate a physics-based model, an autoencoder, and a U-Net-based neural network model as a cohesive multi-fidelity (MF) model for computational FFR assessment. The low-fidelity physics-based model provides computationally efficient, less accurate low-fidelity hemodynamic data. Combined with the features extracted from the autoencoder, those low-fidelity hemodynamic data are fed into the U-Net-based neural network to predict FFR with similar accuracy as high-fidelity hemodynamic simulation, along with a measure of uncertainty. 

The rest of the paper is organized as follows. In section \ref{Methodology}, the proposed AttMulFid is elucidated, focusing on data generation, feature extraction, feature fusion U-Net, uncertainty quantification, and gradient-based attention. Results are presented and discussed in section \ref{Results&Discussion}. Finally, this study concludes with a summary in section \ref{Conclusion}.

\section{Methodology}\label{Methodology}
\begin{figure}[h]
\centering
\includegraphics[width=0.8\textwidth]{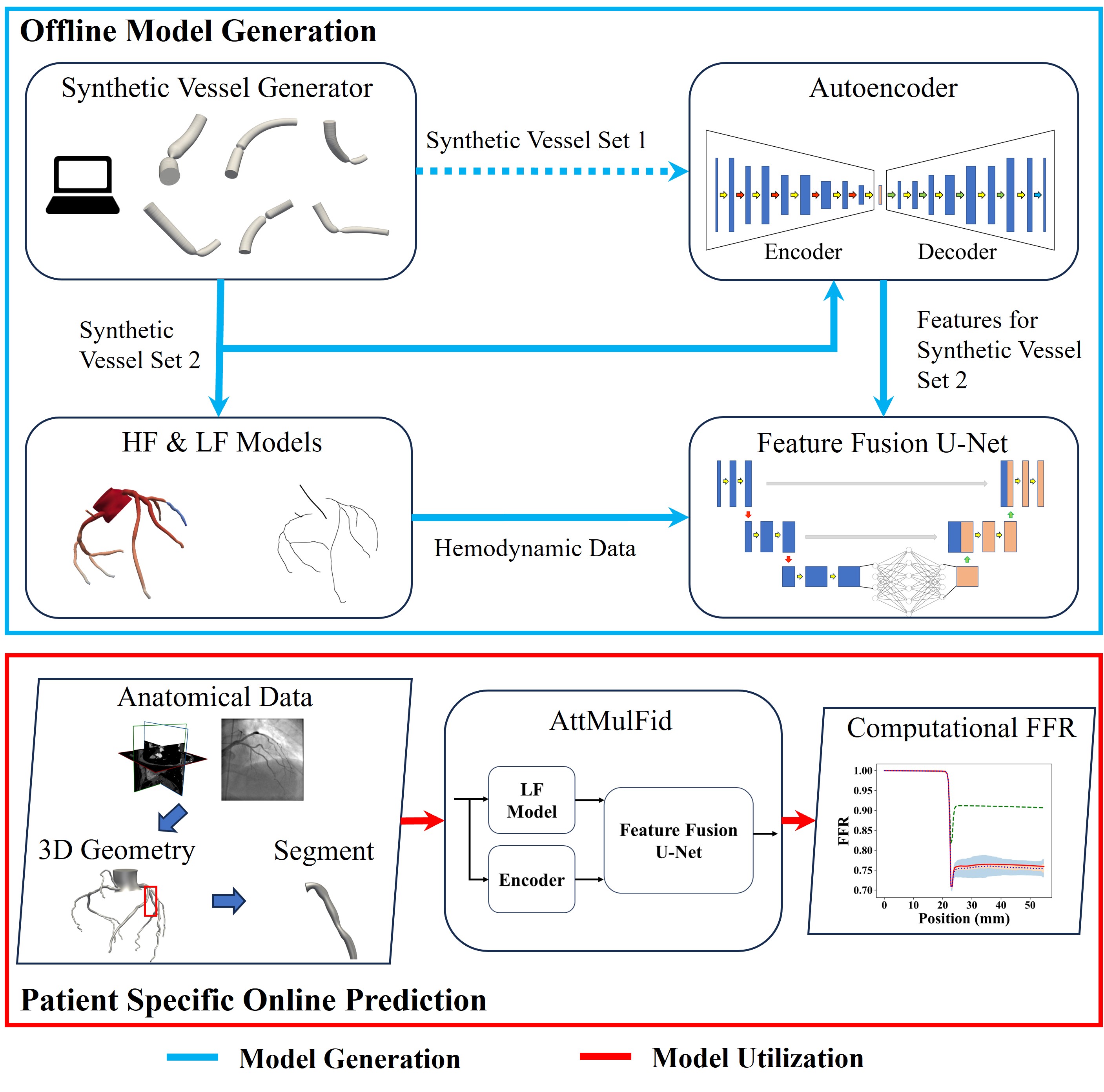}
\caption{The framework of AttMulFid generation and utilization}
\label{fig:Framework}
\end{figure}

This section presents the proposed framework for constructing an AttMulFid and utilizing it for the assessment of FFR. As shown in Fig. \ref{fig:Framework}, it includes two stages: offline model generation and patient-specific online prediction. 

In the offline model generation stage, two sets of synthetic vessel arrays are generated using the synthetic vessel generator first \cite{Iyer2023}. The first set is used to train an autoencoder for feature extraction, focusing on extracting geometric features from the 3D vessel geometry, with additional attention on the key areas, such as stenosis. The second set of vessel arrays is used to generate corresponding hemodynamic data using both the high-fidelity (HF) computational fluid dynamic (CFD) model and the 1D low-fidelity (LF) model. Additionally, the second set of vessel arrays is also fed into the pre-trained autoencoder to obtain lower-rank representations, i.e. geometric features. These hemodynamic data and geometric features are then used to construct a U-Net-based multi-fidelity model with feature fusion (FFU-Net). Specifically, FFU-Net takes the LF hemodynamic data and geometric features as inputs, and predicts HF hemodynamic data. Finally, the LF model for efficient hemodynamic data generation, the encoder of the autoencoder for feature extraction, and FFU-Net for fidelity mapping together form the AttMulFid for FFR assessment. 

In the patient-specific online prediction stage, the anatomical data such as CT image or angiography, is used to reconstruct the 3D geometry \cite{Iyer2023}. Then, the vessel with stenosis is identified for computational FFR assessment. Next, the vessel segment, along with the boundary conditions extracted from anatomical data, is supplied to AttMulFid to predict the velocity and pressure (corresponding to FFR) along the vessel with similar accuracy as HF CFD simulation. First, the boundary conditions and the vessel geometry are supplied to the 1D low-fidelity model to generate computationally efficient low-fidelity hemodynamic data. Simultaneously, the vessel geometry is also given to the encoder part of the autoencoder to extract geometry features. Finally, The FFU-Net utilizes both low-fidelity hemodynamic data, boundary conditions, and geometric features to predict hemodynamic data with a level of accuracy similar to that of HF CFD simulation. It should be noted that the low-fidelity hemodynamic data is entered into the encoder to generate feature maps for low-fidelity hemodynamic data, while boundary conditions and geometric features are directly injected into the latent space in conjunction with the feature maps. This feature injection allows an enhancement in the predicted accuracy of the proposed framework \cite{yang2023neural}.


\subsection{Data Generation}

The synthetic vessel generator, proposed by our previous study \cite{Iyer2023}, is utilized to generate arrays of vessel geometry denoted as $\mathrm{K}\in\mathbb{R} ^{4\times m}$, where the first dimension includes three spacial coordinates of the centerline and corresponding radius, \textit{m} is the sampling resolution along the vessel. The design of the synthetic vessel generator is briefly described in Appendix \ref{synthetic_vessel_generator}, and the full details are presented elsewhere \cite{Iyer2023}. In order to cover the entire design space, a wide range of vessel geometries are produced. Those geometries are characterized by six design variables, as shown in Table \ref{tab:DVs}, including normalized stenosis position, stenosis length, stenosis severity, the taper of the vessel, the maximum radius of the vessel, and the length of the vessel. Additionally, the centerline of the vessel also exhibits random variations within the physiological range of coronary geometries of ten patients.

\begin{table}[h]
\centering
\small
\caption{Design variables and corresponding ranges}
\label{tab:DVs}
\begin{tabular}{c c c}
\hline
Design Variable & Definition & Range \\ \hline
normalized stenosis position    & $x_p$    & $x_p \in [0,1]$    \\ 
stenosis length    & $l_s$    & $l_s \in [0.005,0.03] m$    \\ 
stenosis severity    & $S_s = (D_n-D_s)/D_n\times 100\% $    & $S_s \in [30\%,85\%]$    \\ 
proximal radius    & $r_p $    & $r_p \in [0.0015,0.002] m$    \\ 
taper    & $T_r = r_d/r_p $    & $T_r \in [0.59,0.8]$    \\ 
vessel length    & $l_v$    & $l_v \in [0.04,0.07] m$    \\
centerline curvature    & $\kappa$    & $\kappa \in $ 10 patient data  \\
volume flowrate    & $Q$    & $Q \in [400,1400] mm^3/s$    \\
\hline
\end{tabular}
\end{table}

Physics-based simulations are performed to produce hemodynamic data for the construction of AttMulFid, based on the vessel geometries and boundary conditions. At the inlet of the vessel, a user-prescribed pressure $p_{in} = $ 13065.6 Pa is applied. Additionally, the volume flow rate is considered as another variable to capture various flow behaviors in different branches of coronary arteries. The range of volume flow rates is set between 3.92E-7 and 1.40E-6 $\mathrm{m}^3/\mathrm{s}$. Two computational models are employed to generate hemodynamic data of different accuracies with distinct computational costs, i.e. low-fidelity physics-based model and high-fidelity CFD model, which are elaborated below. In this study, the fluid used for all the simulations is blood, with a viscosity of $\mu$ = 0.004 Pa·s and a density of $\rho$ = 1060 $\mathrm{kg/m}^3$.

\subsubsection{Low-fidelity Physics-based Model}
The low-fidelity physics-based model utilized in this study is derived by integrating the incompressible Navier–Stokes equations over a circular cross-section \cite{xiao2014systematic,alastruey2012arterial}, governing by,
\begin{equation}
\begin{aligned}
&    \left\{
\begin{aligned}
    &\frac{\partial (AU) }{\partial x} = 0 \\
    &U\frac{\partial U }{\partial x}+\frac{1}{\rho}\frac{\partial P }{\partial x} = \frac{f}{\rho A}
\end{aligned}
\right. \quad\text{on}\quad \Omega^{1D},\\
&AU = Q \quad \text{on} \quad \partial\Omega^{1D}_{inlet} \text{, }\partial\Omega^{1D}_{outlet}, \\
&p = p_{in} \quad \text{on} \quad \partial\Omega^{1D}_{inlet}
\end{aligned}
\end{equation}
where $\Omega^{1D}$ is 1D domain of the vessel, \textit{x} is the axial coordinate along the vessel, $A(x)$ is the cross-sectional area of the lumen, $U(x)$ is the cross-sectional average velocity, $P(x)$ is the cross-sectional average pressure, and $f(x)$ is frictional force. The velocity profile is assumed to exhibit axisymmetric characteristics and remain constant along the vessel. It can be mathematically expressed as follows \cite{smith2002anatomically},

\begin{equation}
u(x,\varepsilon)=U(x)\frac{\xi+2}{\xi}\bigg[1-\bigg(\frac{\varepsilon}{r(x)}\bigg)^{\xi}\bigg]
\end{equation}
where $r(x)$ is the lumen radius, $\varepsilon$ is the radial coordinate, and the $\xi$ is the polynomial order. The friction force can be determined from the velocity profile, expressed as $f(x) = -2\mu\pi r\frac{\partial u}{\partial \varepsilon}|_{\varepsilon=r}$. In this study, $\xi = 2$ is utilized, leading to a Poiseuille flow resistance $f = -8\mu \pi U$. The low-fidelity model provides a relatively simplified representation of the hemodynamic behavior and reduces computational complexity and runtime (around 0.001 s per simulation). Although it may not capture all the intricate details of the flow phenomena, it offers faster simulations and requires less computational resources compared to the high-fidelity CFD model.

\subsubsection{High-fidelity Computational Fluid Dynamic Model}

On the other hand, the high-fidelity CFD model offers a more detailed and accurate representation of the hemodynamics. It employs the finite volume method to solve the underlying 3D incompressible Navier–Stokes equations within the vessel,
\begin{equation}
\begin{aligned}
&   \left\{
\begin{aligned}
&\nabla \cdot \mathbf{u} = 0 \\
&\mathbf{u} \cdot \nabla \mathbf{u} = -\frac{1}{\rho} \nabla p + \nu \nabla^2 \mathbf{u}
\end{aligned}
\right. \quad\text{on}\quad \Omega^{3D},\\
&\int_{\partial\Omega^{3D}_{inlet}}{\mathbf{u}} = Q,\\
&\int_{\partial\Omega^{3D}_{outlet}}{\mathbf{u}} = Q,\\
&p = p_{in} \quad \text{on} \quad \partial\Omega^{3D}_{inlet}
\end{aligned}
\end{equation}
where $\mathbf{u}$ is the velocity vector field, \textit{p} is the pressure field, $\rho$ is the fluid density, $\nu$ is the kinematic viscosity. Following a mesh-independence analysis, the computational domain of the vessel is discretized into approximately 1,000,000 structured cells. OpenFOAM software is used to obtain numerical solutions for the governing equations within this discretized domain. However, due to its increased level of complexity, it requires significantly more computational power and time. Each simulation takes approximately two hours to generate the hemodynamic data using a single computing core.

\begin{figure}[h]
\centering
\includegraphics[width=0.4\textwidth]{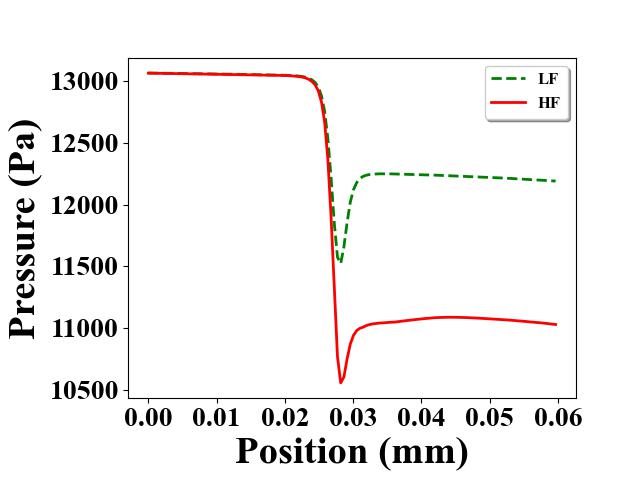}
\caption{Comparison between pressure profiles generated by low-fidelity and high-fidelity with the same vessel geometry and boundary conditions}
\label{fig: HF_LF_comparison}
\end{figure}

Fig. \ref{fig: HF_LF_comparison} demonstrates the comparison between pressure profiles generated by low-fidelity and high-fidelity models given the same vessel geometry and boundary conditions. The pressure profiles exhibit similarity in the proximal region of the vessel. However, notable differences arise after the stenosis due to the breakdown of the low-fidelity model, as it assumes Poiseuille flow resistance, which is no longer valid once the flow passes through the stenosis. In summary, these two models provide a trade-off between accuracy and computational efficiency, and they are capable of generating hemodynamic data with varying levels of fidelity, accuracy, and computational costs for AttMulFid construction.

\subsection{Feature Extraction}

\begin{figure}[h]
\centering
\includegraphics[width=0.8\textwidth]{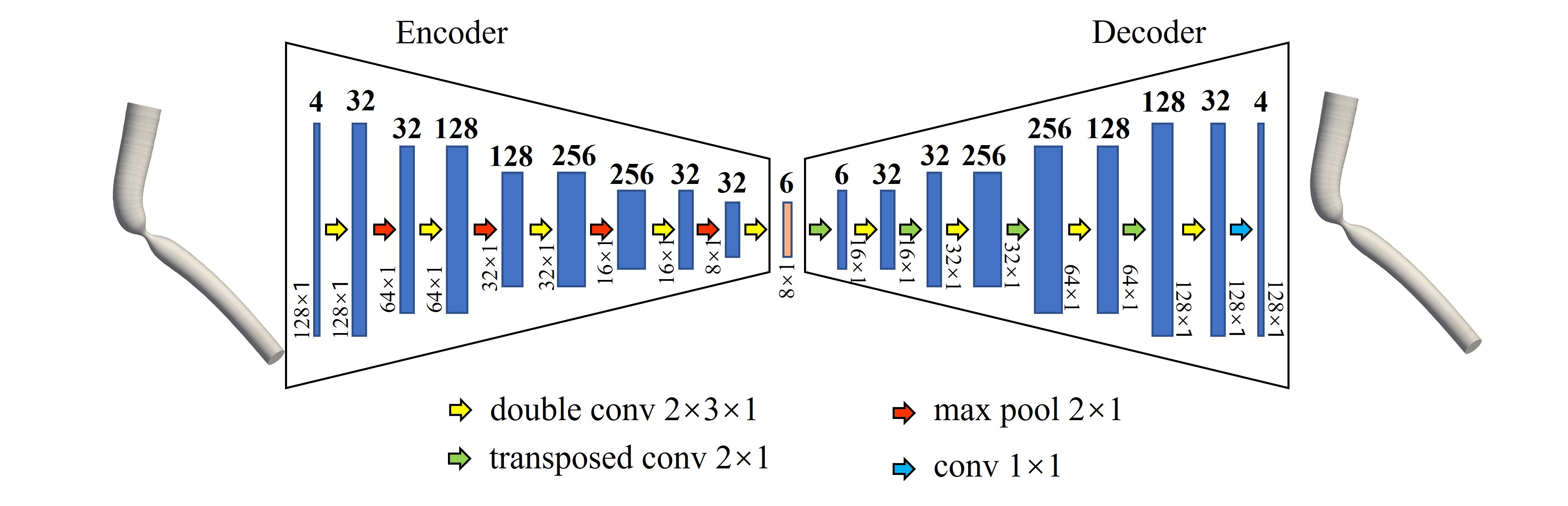}
\caption{The architecture of autoencoder}
\label{fig:AE}
\end{figure}

Autoencoder is designed for extracting important features from the geometry of coronary vessels. It is trained using geometry data of the coronary artery, which serves as both input and output of the model. The geometry of the coronary vessel can be represented by $m\times4$, where \textit{m} denotes the number of samples along the vessel and \textit{m} = 128 is adopted in this study. The second dimension includes three spatial coordinates and radii.  As such, the vessel is assumed to be axisymmetric. The autoencoder comprises two main components: an encoder and a decoder, as shown in Fig. \ref{fig:AE}. The encoder compresses the geometric information into latent space features for dimension reduction and feature extraction. It consists of four pooling layers, each of which is followed by two convolution layers with ReLU activation. The convolution layer, with a kernel size of 2×1 and a stride of 1, compresses the input geometry to generate feature maps. The maximum pooling layer, with a 2×1 kernel and a stride of 2, reduces the dimensions of the feature maps. On the other hand, the decoder performs an inverse process of the encoder. It takes the features from the latent space and reconstructs the geometry of the coronary artery. The decoder consists of four transposed convolution layers, each followed by two convolution layers with ReLU activation. The transposed convolution is an upsampling process, with a 2×1 kernel and a stride of 2. The number of kernels applied in each layer, denoted on the top of the block, corresponds to the number of channels. The detailed architecture of the autoencoder is listed in Table \ref{tab:DAA}. Furthermore, in order to place more attention on key features of the geometry affecting FFR, such as stenosis, the loss function used in the training process is defined as,
\begin{equation}\label{loss}
    L_{autoencoder}=\left\|  \omega\left| y-\hat{y}  \right|  \right\|
\end{equation}
\begin{equation}
    \omega =  (3-r)^7+1
\end{equation}
where y and $\hat{y}$ are original and reconstructed geometry, and $\omega$ are weighting parameters, which is related to the normalized radius (\emph{r}) of the vessel. After the Autoencoder is well-trained, the encoder component of the Autoencoder is employed to compress the geometry of coronary vessels into features in the latent space. These latent space features serve as part of the inputs for the FFU-Net to facilitate fidelity mapping.

\subsection{ U-Net-based Multi-fidelity Model with Feature Fusion}

\begin{figure}[h]
\centering
\includegraphics[width=1\textwidth]{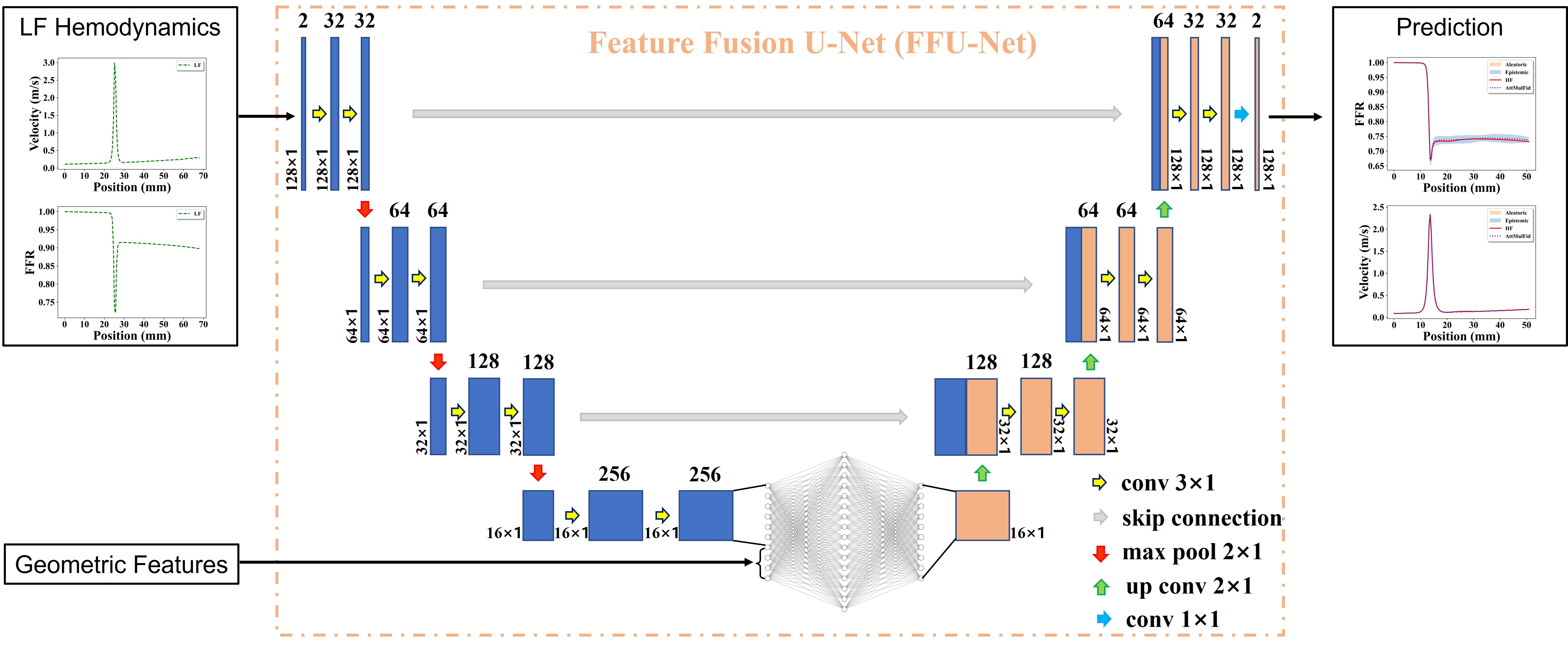}
\caption{Archetechture of FFU-Net}
\label{fig:FFU-Net}
\end{figure}

FFU-Net is designed to map the relationship between low-fidelity and high-fidelity hemodynamic data while integrating boundary conditions and geometric features into the latent space. It is composed of an encoder, a fully connected neural network, and a decoder with skip connections between the encoder and decoder, as depicted in Fig. \ref{fig:FFU-Net}. Within the encoder, a series of convolution layers with ReLU activation and pooling layers, serve the purpose of compressing low-fidelity hemodynamic data into low-fidelity feature maps in the latent space. The fully connected neural network takes on the task of associating low-fidelity hemodynamic feature maps with high-fidelity counterparts. Notably, this network incorporates boundary conditions and geometric features as part of inputs, enhancing its ability to capture the relationship between the two fidelity levels. It has three layers, with 2098, 2048, and 2048 neurons, respectively. The decoder can be regarded as an inverse process of the encoder, reconstructing the high-fidelity hemodynamic data based on high-fidelity feature maps through convolution and transposed convolution with ReLU activation function. In the figure, the yellow arrow represents the convolution with a kernel of 2×1 and a stride of 1 for extracting features. The red arrow denotes the maximum pooling layer with a kernel of 2×1 and a stride of 2, which effectively compresses the feature maps. Conversely, the green arrow signifies the transposed convolution with a kernel of 2×1 and a stride of 2, aiming to expand the dimensions of the feature maps. The number above each block denotes the number of channels within the feature maps, which equivalently corresponds to the number of convolution kernels employed. Additionally, the size of the feature map is indicated beside each block. The details of FFU-Net architecture are elaborated in Table \ref{tab:DFFU}. In summary, the encoder compresses low-fidelity hemodynamic data into low-fidelity feature maps. Subsequently, these low-fidelity feature maps, alongside boundary conditions and geometric features, navigate through the fully connected neural network, yielding high-fidelity feature maps. Finally, the decoder leverages these high-fidelity feature maps to predict hemodynamic data with an accuracy matching that of high-fidelity hemodynamic data.

\subsection{Uncertainty Quantification}

As mentioned above, the AttMulFid consists of three distinct elements: the 1D model, encoder, and FFU-Net. The 1D model is a deterministic physics-based model that computes low-fidelity hemodynamic information without any inherent uncertainty. In contrast, both the encoder and FFU-Net are data-driven models trained using input-output data pairs, introducing uncertainties in their predictions. The autoencoder (encoder) is trained using a large number of synthetic vessels as both input and output, therefore uncertainty existing in the autoencoder is relatively small and omitted.

Two types of uncertainty are quantified for FFU-Net: aleatoric uncertainty and epistemic uncertainty \cite{hullermeier2021aleatoric, kendall2017uncertainties, abdar2021review}. Aleatoric uncertainty captures the inherent noise in the data due to the non-deterministic nature of the input-output relationship. Importantly, this type of uncertainty is irreducible, regardless of the quantity of training data. Within aleatoric uncertainty, two subcategories are distinguished: homoscedastic aleatoric uncertainty and heteroscedastic aleatoric uncertainty. Homoscedastic aleatoric uncertainty remains consistent for each input, whereas heteroscedastic aleatoric uncertainty varies across different model inputs. In this study, we make the simplifying assumption that aleatoric uncertainty is homoscedastic. This means that the aleatoric uncertainty is assumed to be constant across all inputs. This type of uncertainty is quantified using the Gaussian negative log-likelihood loss function \cite{valdenegro2022deeper, zhang2023risk}:

\begin{equation}
\text{NLL} = \frac{(y-\hat{y})^2}{2\sigma^2_{\text{aleatoric}}} + \frac{1}{2}\log(\sigma_{\text{aleatoric}})
\end{equation}
where $y$ is the true target value, $\hat{y}$ is the predicted value and $\sigma^2$ is the variance of the Gaussian distribution.

On the other hand, epistemic uncertainty arises from a lack of knowledge and refers to ignorance of the agent or decision maker \cite{abdar2021review}. This type of uncertainty can be reduced as more data becomes available \cite{kendall2017uncertainties}. Estimating the epistemic uncertainty of a neural network requires computing the variance of the model prediction. Various methods were developed to estimate epistemic uncertainty, such as Bayesian neural networks, Monte Carlo dropout models, and deep ensemble models \cite{hullermeier2021aleatoric,kendall2017uncertainties,lakshminarayanan2017simple,gal2016dropout,ganaie2022ensemble}. In this study, deep ensemble method is employed to quantify the epistemic uncertainty of FFU-Net. Specifically, an ensemble of $M = 5$ FFU-Net models is constructed explicitly and statistical variance is calculated among their predictions. Mathematically, epistemic uncertainty is defined as:
\begin{equation}
\sigma^2_{\text{epistemic}} = \frac{1}{M} \sum_{j=1}^{M}(y_{j}^{2}-\hat{y}_{j}^{2})
\end{equation}
The total uncertainty for the prediction is computed by combining the aleatoric and epistemic uncertainties through the statistical law of total variance:
\begin{equation}
\sigma^2_{\text{total}} = \sigma^2_{\text{aleatoric}}+\sigma^2_{\text{epistemic}}
\end{equation}

\subsection{Gradient-based Attention}

\begin{figure}[h]
\centering
\includegraphics[width=0.7\textwidth]{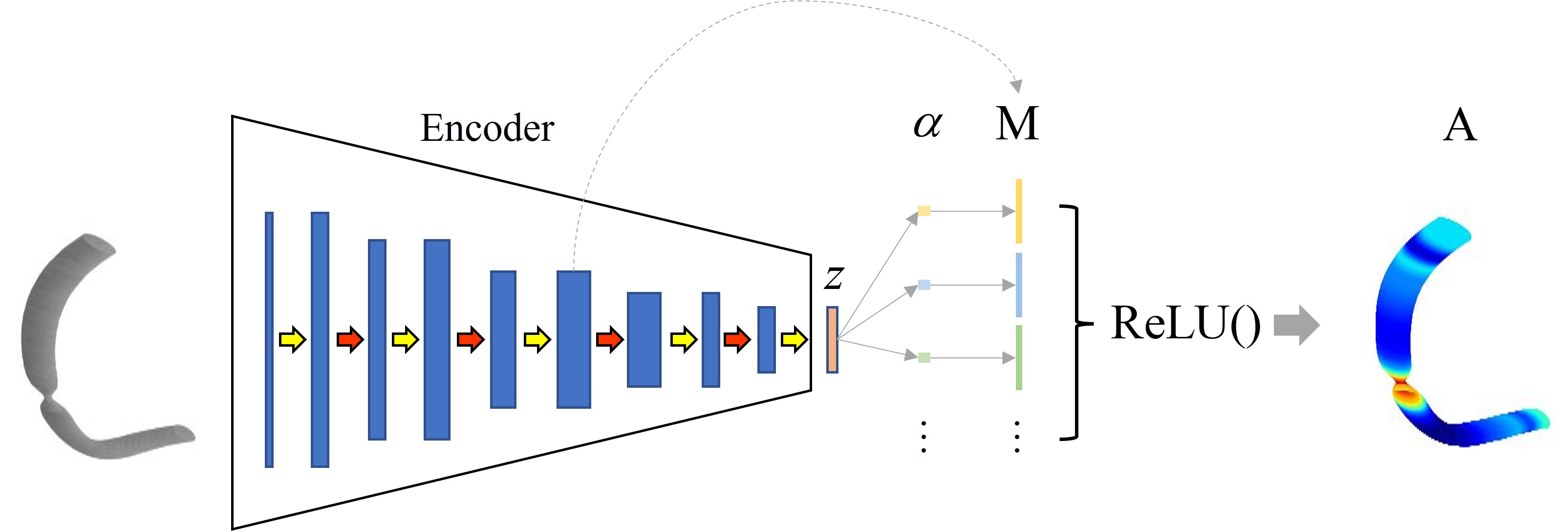}
\caption{Gradient-based attention generation}

\label{fig:GA_Flowchart}
\end{figure}

Other than robustness and performance assurances, interpretability is a also crucial requirement for machine learning models, particularly in safety-critical and consumer-focusing domains like healthcare, where trust in deep learning systems is paramount for safety and consumer confidence \cite{hassanin2022visual,selvaraju2017grad,zhang2021survey}. In this section, gradient-based attention is introduced to provide a visual explanation of neural network learning using attention maps \cite{liu2020towards}. As shown in Fig. \ref{fig:GA_Flowchart}, given the feature vector \textbf{z} obtained in the latent space of the neural network, backpropagation is applied to compute the gradient of most comprehensive feature maps $\textbf{M}\subseteq \mathbb{R}^{n\times l}$ with respect to the mean of \textbf{z}. This calculation results in coefficient $\alpha_{k}$, each corresponding to a specific feature map $\textbf{M}_{k}$. Subsequently, the gradient-based attention $\textbf{A}$ is computed by applying a ReLU function to the linear combination of these feature maps $\textbf{M}_{k}$ and corresponding coefficients $\alpha_{k}$. Mathematically, this process can be expressed as follows:
\begin{equation} \label{Eq:Attention}
    \textbf{A} = \textrm{ReLU}(\sum_{k=1}^{n}\alpha_{k}\textbf{M}_{k} )
\end{equation}
\begin{equation}
    \alpha_{k} =\frac{1}{l}\sum_{i=1}^{l}(\frac{\partial \bar{\textbf{z}}}{\partial \textbf{M}_{k}^{i}})
\end{equation}
where \emph{n} is the number of feature maps, \emph{l} is the length of each feature map. By analyzing the attention map, the mechanism of neural network learning can be visualized and explained.

\section{Results and Discussion}\label{Results&Discussion}

In this section, the results are presented to demonstrate the feasibility of the framework in terms of feature extraction from the vessel geometry and hemodynamic assessment. In addition to synthetic data, a cohort of three patient data with extensive cardiovascular disease testing is collected to evaluate the performance of the proposed framework. These patients exhibit stable epicardial disease in the coronary artery, and CT images, angiograms, and invasive FFR data are collected for validation purposes.

\subsection{Feature Extraction and Geometry Reconstruction} 

The autoencoder is used for vessel geometry reconstruction with the encoder component specifically designed for feature extraction. To validate that the extracted features encompass all the essential information present in the original geometry, the reconstructed vessel geometries are compared with their corresponding originals. During the training process, a dataset consisting of 100,000 synthetic vessel segments is collected. This dataset is generated using Latin Hypercube Sampling (LHS) and is partitioned as follows: 80\% of the samples are utilized for training the model, 10\% serve as validation data for monitoring and fine-tuning the model during training, and the remaining 10\% are reserved for testing the model's performance post-training. Additionally, three vessel segments extracted from patients' CT scans are utilized to further validate the performance of our feature extraction method. In total, 48 geometric features are extracted from the latent space. It's worth noting that the original input geometry is represented as a $4\times128$ matrix. Consequently, our feature extraction results in a remarkable reduction of dimensionality, reducing 90.6\% of dimensions originally present in the input geometry.

\subsubsection{Synthetic Vessels}

\begin{figure}[h]
\centering
\includegraphics[width=1\textwidth]{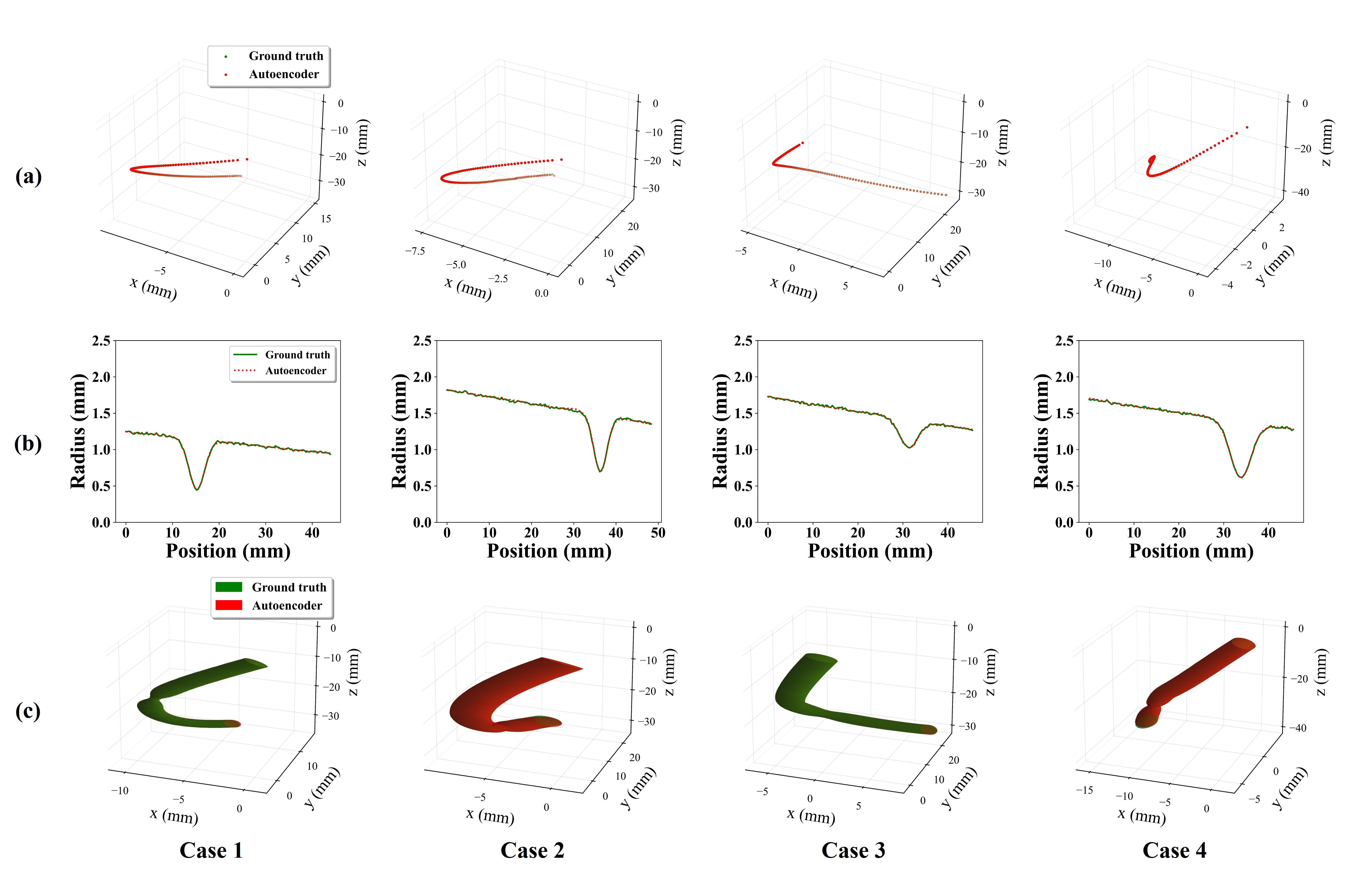}
\caption{Geometry reconstruction for synthetic vessels: (a) centerline, (b) radius, and (c) 3D geometry.}
\label{fig:Synthetic_Geo_Rec}
\end{figure}

First, feature extraction is applied to synthetic vessels in the testing data set to illustrate the reconstruction performance, as shown in Fig. \ref{fig:Synthetic_Geo_Rec}. The rows from top to bottom display the centerline, radius, and 3D geometry comparisons between ground truth and reconstruction geometries. Results clearly illustrate that the reconstructed geometries exhibit an outstanding level of agreement with the ground truth geometries with negligible discrepancies. The mean square error (MSE) calculated across the entire testing set is 2.16E-9. This MSE value strongly indicates that the low-dimensional geometric features extracted in the latent space are fully capable of representing the entire geometric information present in the original data.

\begin{figure}[h]
\centering
\includegraphics[width=1\textwidth]{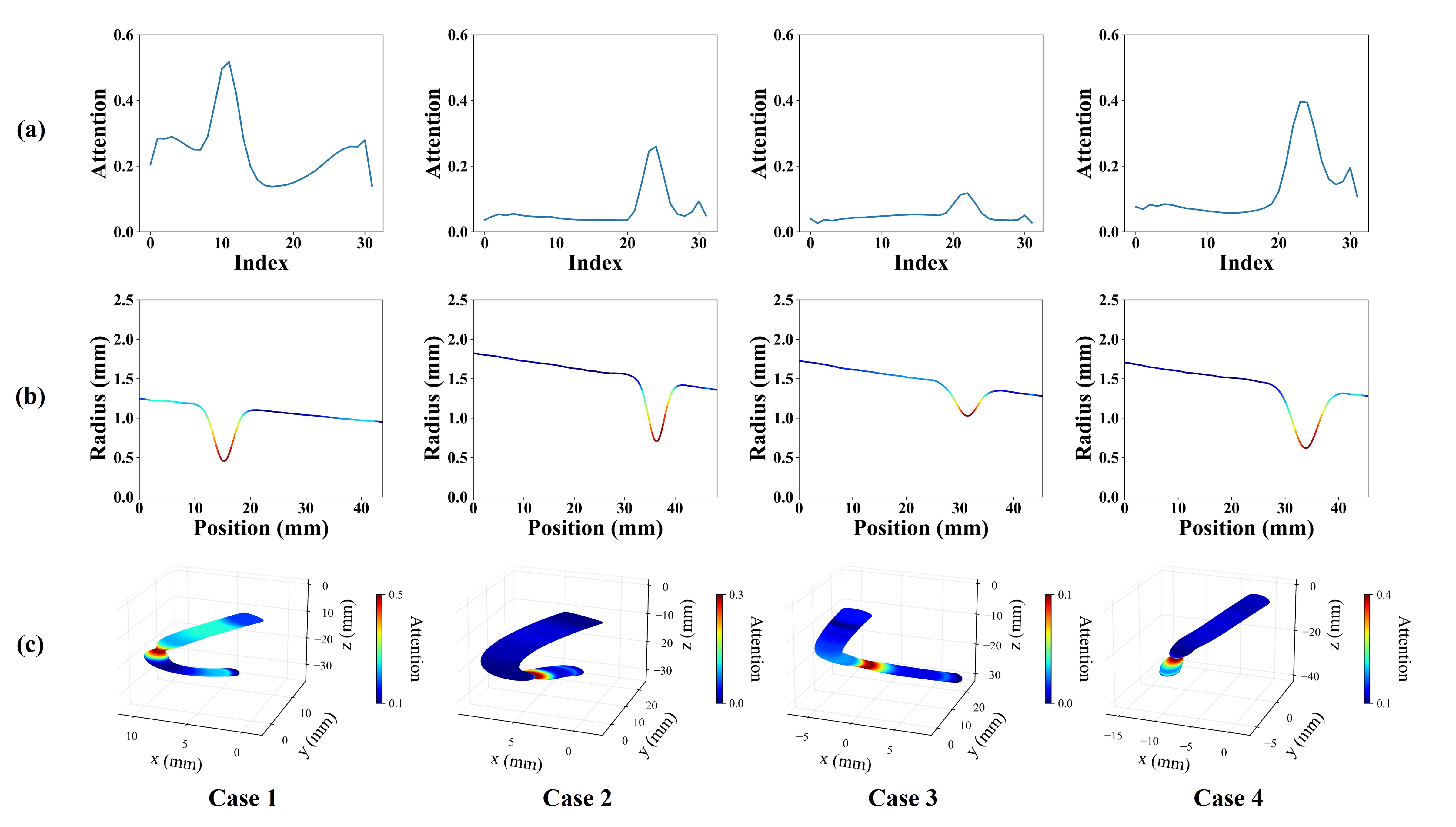}
\caption{Gradient-based attention for synthetic vessels: (a) magnitude, (b) projection on the radius, and (c) projection on 3D geometry.}
\label{fig:Synthetic_GA}
\end{figure}

To delve into the learning process of the autoencoder and provide a visual explanation, the gradient-based attentions are presented for four testing cases, as shown in Fig. \ref{fig:Synthetic_GA}. The top row represents the attention maps computed in Eq. \ref{Eq:Attention}, while the middle and bottom rows illustrate the projection of the attention maps onto the radius and 3D geometry, respectively. There are several interesting observations emerged from these attention maps. First, the attention is notably higher in regions where stenosis is present compared to other areas. This observation demonstrates that the extracted features effectively identify stenosis and allocate a greater degree of attention to such regions. Furthermore, the attention maps also encapsulate specific properties of stenosis, such as stenosis length (comparing case 2 and case 4), proximal radius (comparing case 1 and case 2), and the degree of stenosis (comparing case 3 and case 4). Longer stenosis lengths, smaller proximal radii, and higher degrees of stenosis correspond to higher magnitudes of attention in the maps. In summary, the geometric features learned by the autoencoder exhibit the capability to identify and intelligently allocate attention based on crucial properties of stenosis, thereby providing valuable insights into these key characteristics.

\subsubsection{Patient Vessels}

The patient vessels are extracted from CT images using CRIMSON (CardiovasculaR Integrated Modeling and SimulatiON) open-source software \cite{arthurs2021crimson}. These patient vessels are specifically derived from the right coronary artery, making them more closely aligned with our training data. In contrast, vessels originating from the left coronary artery exhibit significant differences in terms of centerline curvature and radius compared to our training dataset. Consequently, these left coronary artery vessels have been excluded from the validation process for our proposed framework.

\begin{figure}[h]
\centering
\includegraphics[width=1\textwidth]{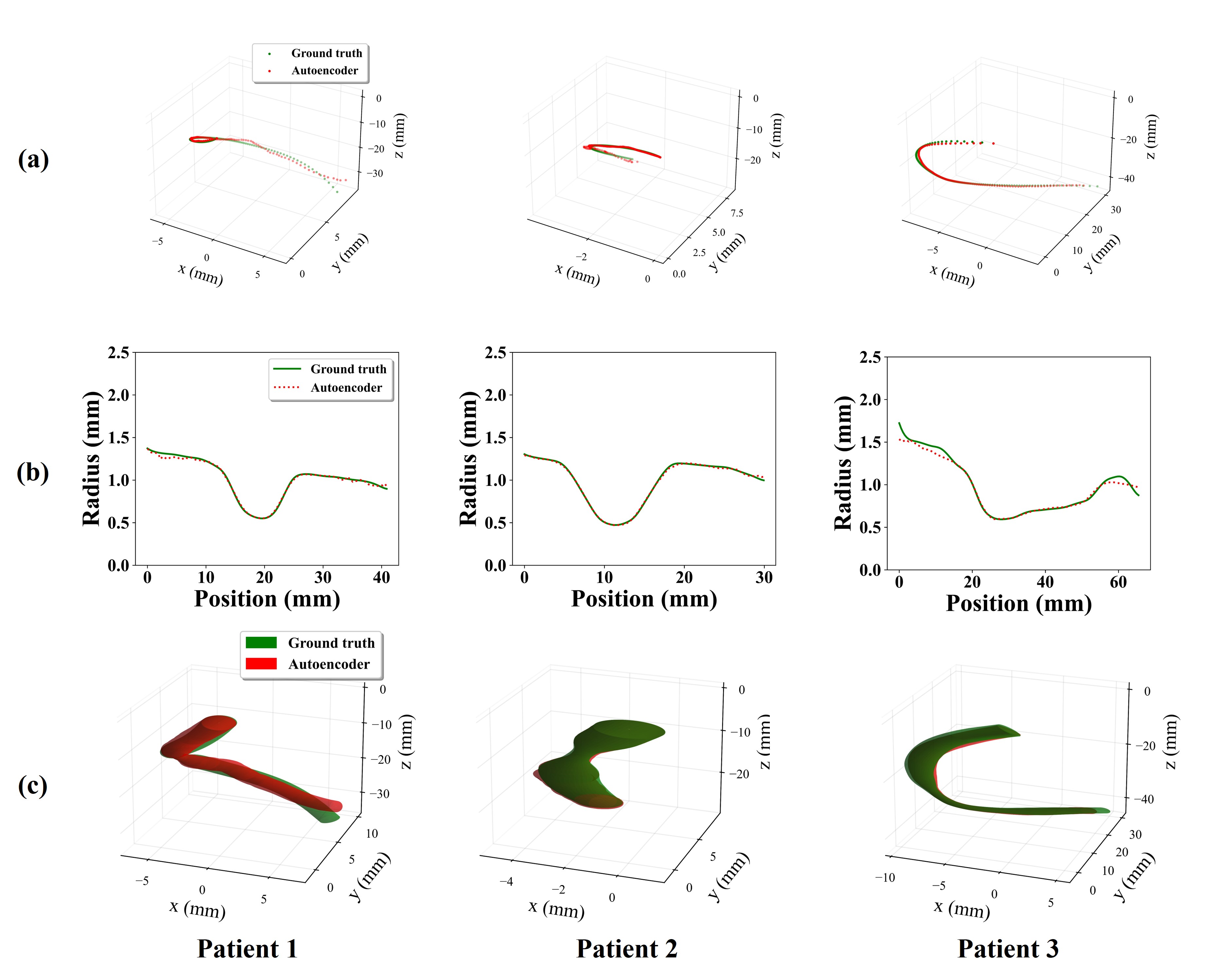}
\caption{Geometry reconstruction for patient vessels: (a) centerline, (b) radius, and (c) 3D geometry.}
\label{fig:Patient_Geo_Rec}
\end{figure}

Same as the synthetic vessels, the patient vessels are also represented as $4\times m$ array, where m is the sampling resolution along the vessel. Then, the patient vessel geometries are ready for evaluation. Fig. \ref{fig:Patient_Geo_Rec} shows the results of geometry reconstruction for three patient vessels. Again, the rows, from top to bottom, display the centerline, radius, and 3D geometry comparisons between ground truth and reconstruction geometries. Results demonstrate that the reconstructed geometries capture the general trend of the patient geometries with small discrepancies. In general, the reconstruction of vessel radius is more accurate compared to the centerline reconstruction. This is primarily due to higher weights to the loss associated with radius in the definition of the loss function, especially in regions with stenosis. The reconstruction MSE of patient vessels is 4.3E-8, which is slightly higher than that of synthetic vessels. This difference can be attributed to the fact that patient vessels exhibit greater complexity, and a higher degree of variability compared to synthetic vessels used to train the proposed model. Therefore, the inference of patient vessels can be considered as extrapolation,  resulting in slightly lower accuracy. Nevertheless, the overall agreements indicate that the autoencoder effectively compresses the geometric information of the vessels into low-dimensional features, while preserving the important geometric details.

\begin{figure}[h]
\centering
\includegraphics[width=1\textwidth]{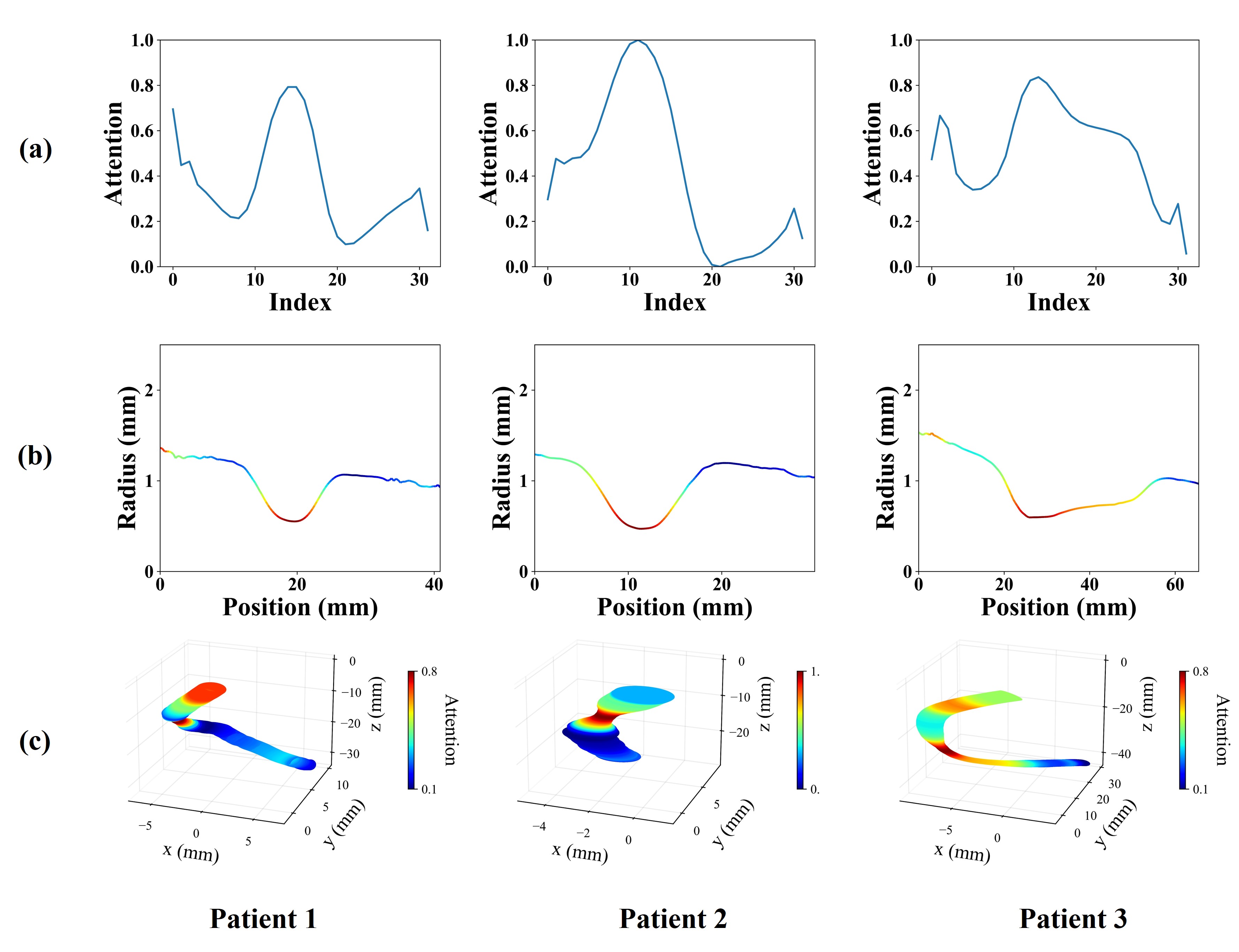}
\caption{Gradient-based attention for patient vessels: (a) magnitude, (b) projection on the radius, and (c) projection on 3D geometry.}
\label{fig:Patient_GA}
\end{figure}

In Fig. \ref{fig:Patient_GA}, the gradient-based attentions are displayed to visually explain the feature extraction process for three patient vessels. The rows from top to bottom illustrate the attention maps, and their projections onto the radius and 3D geometry, respectively. Results demonstrate that geometric features are effective in identifying stenosis and assigning more attention to that region, even though the stenoses in patient vessels may be diffused (Patient 3), and do not conform to the ideal Gaussian curves assumed in the synthetic vessel generator during training. Additionally, the magnitude of attention is correlated with the stenosis severity, with more severe stenosis receiving higher attention. In summary, the autoencoder trained using synthetic vessels maintains its capability to identify and intelligently allocate attention based on stenosis properties in patient-specific vessels.

\subsection{Hemodynamic Assessment} 

\begin{figure}[h]
\centering
\includegraphics[width=0.5\textwidth]{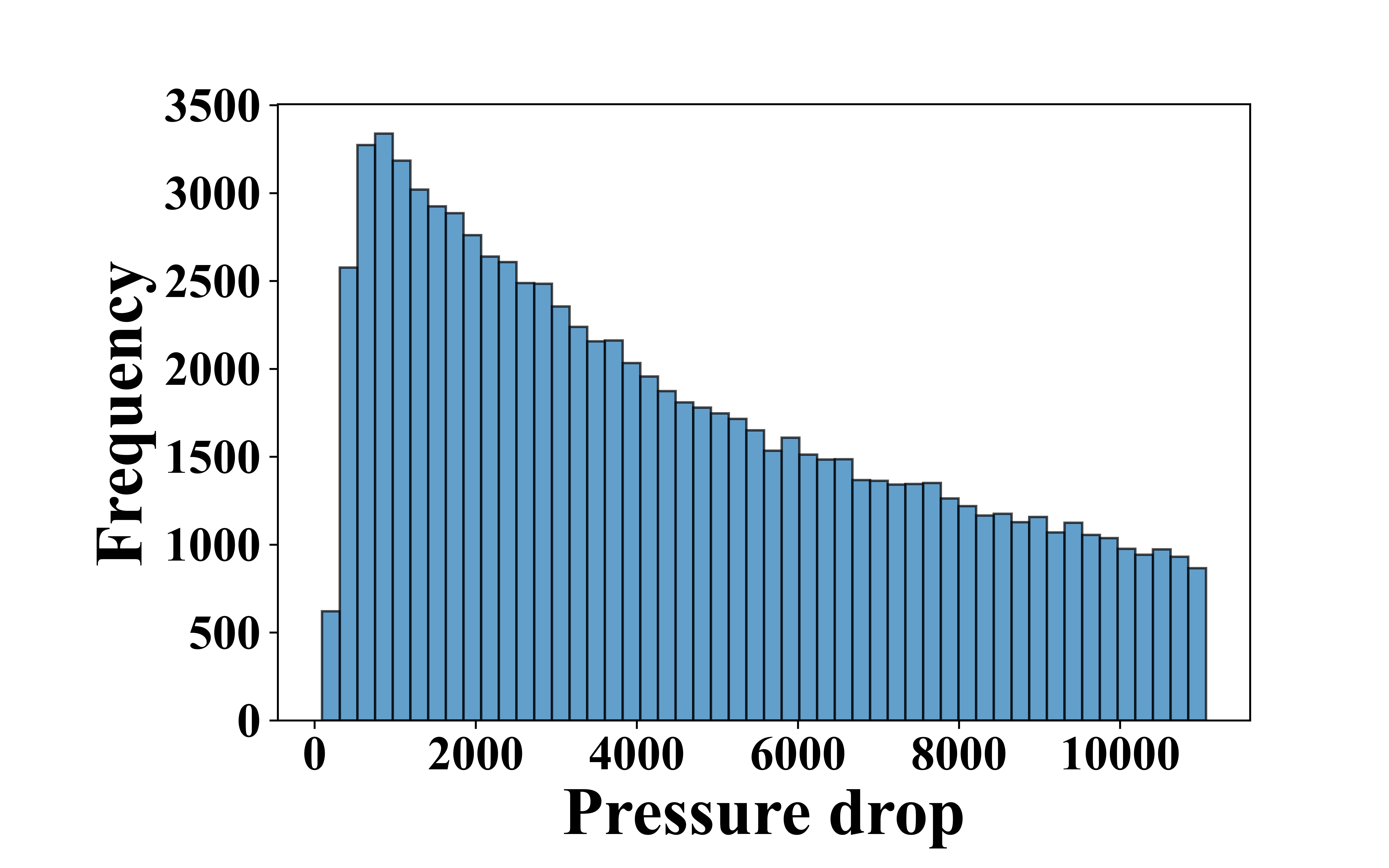}
\caption{Pressure drop distribution for all the training data}
\label{fig:P_drop}
\end{figure}

Following the successful verification of the feature extraction process, the encoder component can be integrated into AttMulFid to extract features from the geometry. These features are subsequently fused into the latent space of FFU-Net. Our study showed that employing LHS with the design variables specified in Table \ref{tab:DVs} introduces a notable bias in the distribution of pressure drops. Specifically, a considerable proportion of hemodynamic simulations end up with relatively lower pressure drops. To enhance the accuracy of the cases characterized by larger pressure drops (i.e., more diseased scenarios), the LHS is performed based on a modified set of variables, including $x_p,l_s, S_s^{10}, r_p^{10}, T_r^{20},l_v^{10}, Q$. Consequently, a relatively uniform distribution of the pressure drops is obtained, as shown in Fig. \ref{fig:P_drop}

Next, those samples are supplied to both high-fidelity and low-fidelity hemodynamic simulations, for generating both high-fidelity and low-fidelity hemodynamic data pairs. Of these data pairs, 95\% are allocated for training the model, while the remaining 5\% are used for validation purposes. Additionally, 1000 samples were generated in the same manner to produce testing data pairs to evaluate the performance of the AttMulFid model.

\subsubsection{Synthetic Vessels}

\begin{figure}[h]
\centering
\includegraphics[width=1\textwidth]{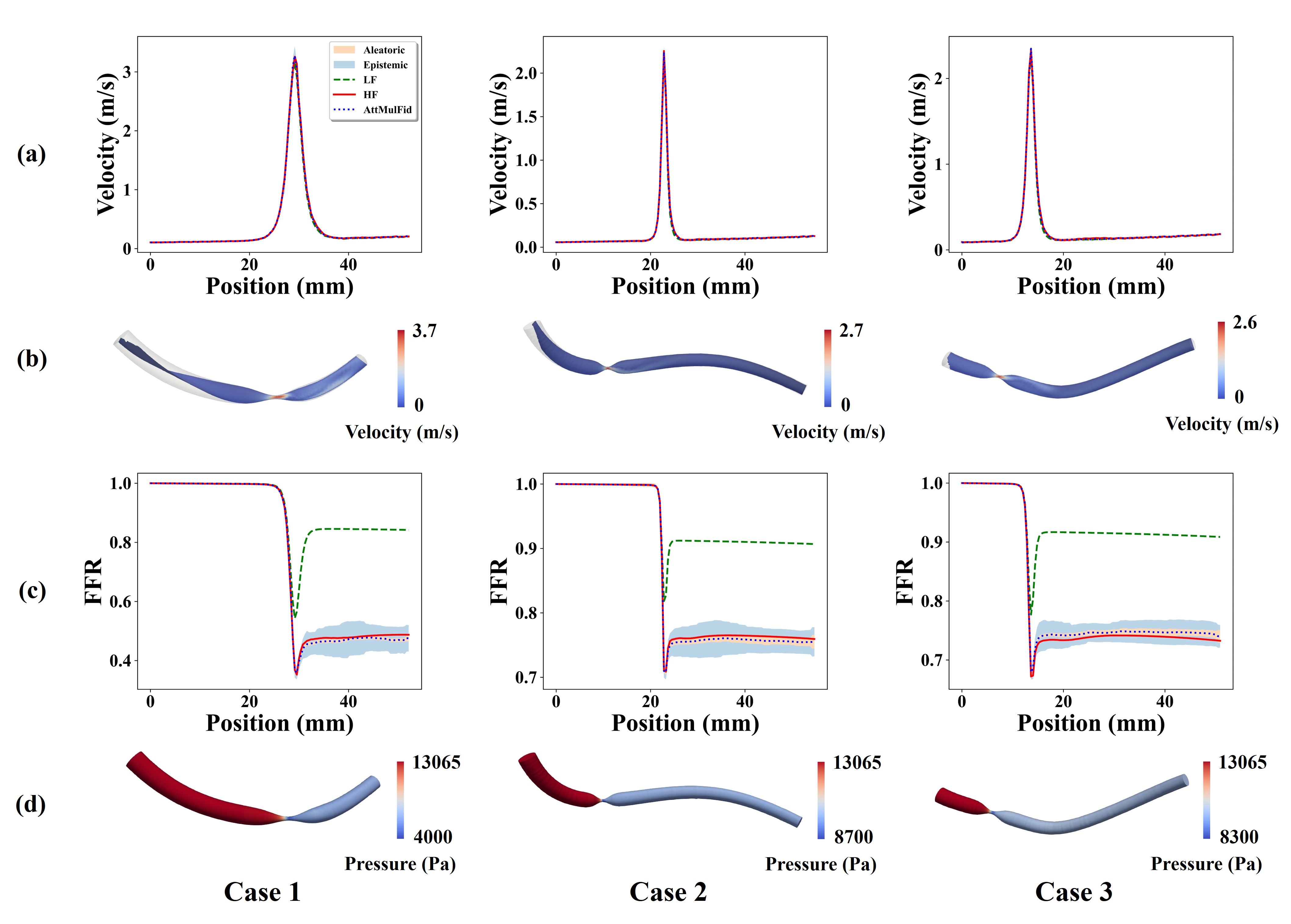}
\caption{Hemodynamic comparison for synthetic vessels: (a) velocity comparison, (b) velocity contour plots of CFD simulations, (c) FFR comparison, (d) pressure contour plots of CFD simulations}
\label{fig:Hemodynamic_synthetic_vesssel}
\end{figure}

Similar to the previous section, the performance of AttMulFid is first evaluated using synthetic vessels. Fig. \ref{fig:Hemodynamic_synthetic_vesssel} illustrates the hemodynamic comparisons among LF, HF simulations, and the predictions generated by AttMulFid for three testing cases within the testing set. Corresponding MSEs between AttMulFid and CFD,  as well as uncertainties of AttMulFid prediction of three testing cases shown in Fig. \ref{fig:Hemodynamic_synthetic_vesssel} and average values across all testing cases, are listed in Table \ref{tab:MSEUQsyn}. The results show that velocity predictions from AttMulFid closely agree with the velocities calculated from the HF simulations, with negligible discrepancies and uncertainty. This agreement is expected, given that the LF simulation also accurately computes velocity profiles. On the other hand, despite substantial discrepancies existing between LF and HF simulations in the FFR (pressure) profiles, AttMulFid remarkably captures the FFR (pressure) profiles with minimal differences (with an average MSE of 1.2E-04 for all testing cases). The HF FFR profiles fall within the uncertainty bounds quantified by the proposed method, indicating that the proposed method can accurately quantify where the HF simulations are most likely located. Compared to aleatoric uncertainty, epistemic uncertainty is the major source of variations in AttMulFid prediction. Notably, the uncertainties predominantly manifest immediately after the stenosis, a phenomenon consistent with the underlying physics, i.e. the LF model becomes less reliable after the stenosis due to the Poiseuille resistance assumption. Moreover, AttMulFid boasts an incredibly short prediction time of approximately 0.002s, which is significantly faster than the two-hour duration required for CFD-based FFR calculations and the approximately half-hour duration of invasive measurements. Consequently, AttMulFid emerges as a promising computational approach for non-invasive, efficient, and accurate FFR assessment.

\begin{table}[h]
\centering
\small
\caption{MSEs and uncertainties for synthetic vessels}
\label{tab:MSEUQsyn}
\begin{tabular}{c c c c c}
\hline
  & Case 1 & Case 2 & Case 3 & Average \\ \hline
MSE & 6.1E-05 & 1.9E-05 & 4.7E-05 & 1.2E-04    \\ 
Aleatoric Uncertainty ($\sigma^2_{\text{aleatoric}}$) & 4.6E-06 & 4.6E-06 & 4.6E-06 & 4.6E-06    \\ 
Epistemic Uncertainty ($\sigma^2_{\text{epistemic}}$) & 2.5E-04 & 6.9E-05 & 7.0E-05 & 6.0E-05    \\ 
Total Uncertainty ($\sigma^2_{\text{total}}$) & 2.6E-04 & 7.3E-05 & 7.4E-05 & 6.5E-05    \\ 
\hline
\end{tabular}
\end{table}

\subsubsection{Patient Vessels}

\begin{figure}[h]
\centering
\includegraphics[width=1\textwidth]{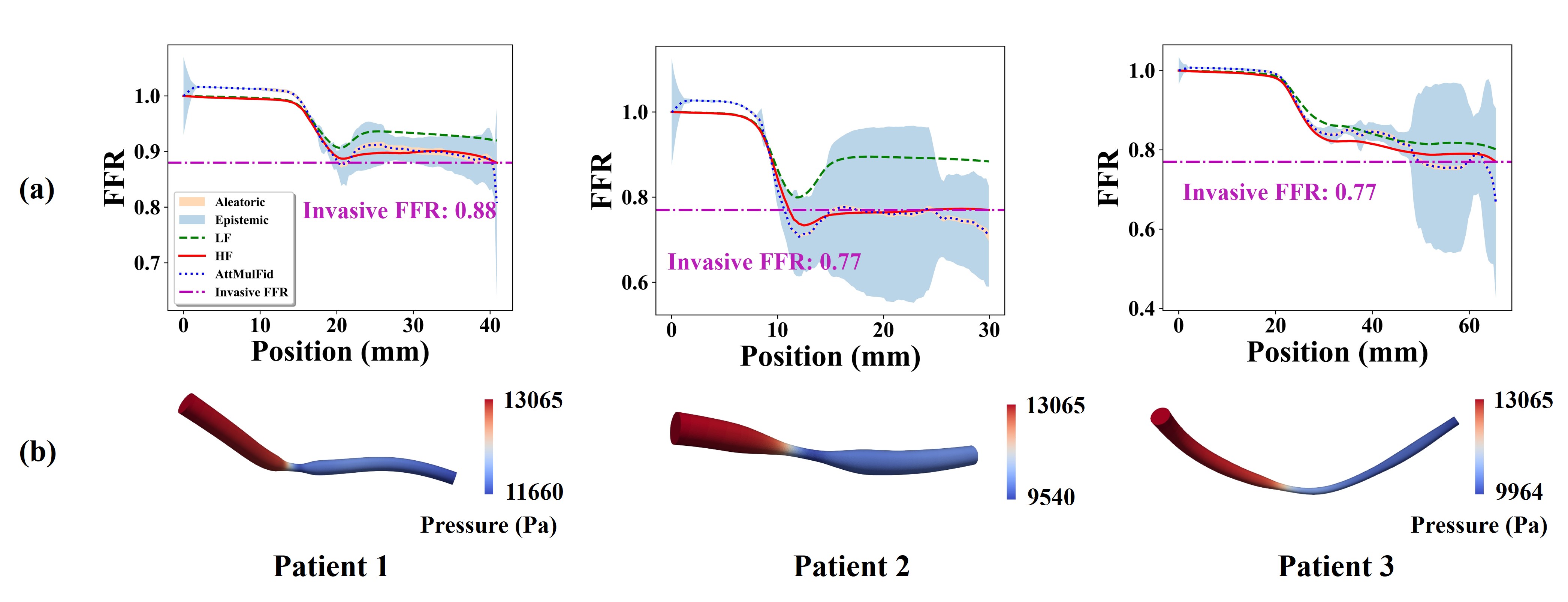}
\caption{Hemodynamic comparison for patient vessels: (a) FFR comparison, (b)  pressure contour plots of CFD simulations}
\label{fig:Hemodynamic_patient_vesssel}
\end{figure}

Ultimately, FFRs predicted by the proposed method are validated by comparing them with invasive FFR measurements obtained through pressure wire for three patient vessels. The flow information is extracted by performing CFD simulation using pressure waveforms acquired through invasive measurement as boundary conditions \cite{morris2021novel}. The FFR predictions for patient-specific vessels using AttMulFid are compared with those computed from LF model, HF CFD model, and invasive measurement, as illustrated in Fig. \ref{fig:Hemodynamic_patient_vesssel}. Similarly, Table \ref{tab:MSEUQspatient} enumerates the MSEs between AttMulFid and CFD, along with the uncertainties in AttMulFid predictions for three patient cases illustrated in Fig. \ref{fig:Hemodynamic_patient_vesssel}. The average values across all patient cases are also included in the table. AttMulFid predictions correct the FFR estimates obtained from the LF model, striving to align with the FFR predictions produced by the CFD model. Nevertheless, noticeable discrepancies (MSE=6.0E-04) can be observed between the predictions of AttMulFid and the CFD model. This discrepancy can be attributed to the greater complexity of patient vessels, which lie beyond the scope of our training data encompassing synthetic vessels. Consequently, the FFR prediction for patient vessels can be considered as extrapolation,  leading to a modest reduction in accuracy. For the same reason, AttMulFid also tends to generate relatively larger uncertainties, especially after stenosis. Compared with the invasive method, the AttMulFid predictions exhibit decent accuracy with 6.7E-03 MSE.

\begin{table}[h]
\centering
\small
\caption{MSEs and uncertainties for patient vessels}
\label{tab:MSEUQspatient}
\begin{tabular}{c c c c c}
\hline
  & Case 1 & Case 2 & Case 3 & Average \\ \hline
MSE & 1.8E-04 & 4.8E-04 & 4.9E-04 & 3.9E-04    \\ 
Aleatoric Uncertainty ($\sigma^2_{\text{aleatoric}}$) & 4.6E-06 & 4.6E-06 & 4.6E-06 & 4.6E-06    \\ 
Epistemic Uncertainty ($\sigma^2_{\text{epistemic}}$) & 2.2E-04 & 3.9E-03 & 2.4E-03 & 2.2E-03    \\ 
Total Uncertainty ($\sigma^2_{\text{total}}$) & 2.2E-04  & 3.9E-03 & 2.4E-03 & 2.2E-03   \\ 
\hline
\end{tabular}
\end{table}

\section{Conclusion}\label{Conclusion}

This paper introduces a novel attention-based multi-fidelity machine learning model (AttMulFid) framework for computationally efficient and accurate FFR assessment with uncertainty measurement. The framework comprises two main stages: offline model generation and patient-specific online prediction. In the offline stage, the AttMulFid is designed to accurately and efficiently predict FFR using data with different fidelities, accuracies, and computational costs. It is composed of a low-fidelity model for efficient but less accurate hemodynamic data generation, an autoencoder for feature extraction, and a U-Net-based multi-fidelity model for mapping data across fidelities. Specifically, the autoencoder is trained first using a set of synthetic vessels to extract geometric features, with additional attention on the key areas, such as stenosis. Subsequently, another set of synthetic vessels is supplied to both high-fidelity and low-fidelity models to generate hemodynamic data at both high and low fidelities. The geometric features for the second set of synthetic vessels are obtained using the pre-trained autoencode. The U-Net-based multi-fidelity model is then trained using both the hemodynamic data and geometric features to predict FFR with a level of accuracy comparable to high-fidelity hemodynamic simulations while also providing an uncertainty measure. In the patient-specific online prediction stage, the verified AttMulFid is employed to replace the CFD model to compute FFR for patient-specific vessels with uncertainty measurement. 
 
Results demonstrate that AttMulFid obtained in the offline model generation stage achieves high accuracy in predicting FFR with minimal uncertainty. It can predict FFR with an MSE of 1.2E-04 relative to CFD and the average total uncertainty is just 6.5E-05 for synthetic vessels. Additionally, when applied to patient-specific vessels, AttMulFid's predictions exhibit an MSE of 3.9E-04 and a total uncertainty of 2.2E-03 relative to CFD. Moreover, AttMulFid predictions show a decent MSE of 6.7E-03 when compared to invasive methods. In summary, AttMulFid represents a viable approach for non-invasive, fast, and accurate FFR assessment, which can significantly facilitate the diagnosis of coronary artery disease. Future research will focus on extending this approach to cover the more complex coronary artery segments and predicting FFR for the entire coronary tree.

\section*{Appendix}

\begin{appendices}
\setcounter{table}{0} 
\renewcommand\thesection{\Alph{section}}
\renewcommand\thesubsection{A.\arabic{subsection}}
\renewcommand\thetable{A\arabic{table}}

\subsection{Synthetic Vessel Generator}\label{synthetic_vessel_generator}
The synthetic vessel generator is designed to produce a large number of synthetic vessels for hemodynamic simulations and the training of the neural network. The procedure for generating the centerlines and radii of the synthetic vessel is outlined as follows: (1) The patient-specific vessel centerline and radii $(x,y,z,r)$ (consisting of 17 control points for each vessel) are extracted from 3D CTA images of ten different patients. (2) For each control point, the Gaussian distribution is formulated with mean $(\bar{x},\bar{y},\bar{z})$ and standard deviation $(\sigma_x,\sigma_y,\sigma_z)$. Additionally, the minimum and maximum vessel radii $(r_{min},r_{max})$ are also determined. (3) New control points are sampled based on uniform distribution within the range $(\bar{x}\pm\sigma_x,\bar{y}\pm\sigma_y,\bar{z}\pm\sigma_z)$. The new centerline is obtained by fitting a spline to these new control points and represent the curve using $m$ points on the fitted curve, resulting in a $3\times m$ matrix. (4) The radius vector decreases linearly along the vessel length within the range of $(r_{min},r_{max})$. The radius vector embedded the information about stenosis, assuming Gaussian profiles. (5) The centerline vector and the radius vector are combined as a $4\times m$ matrix to represent the synthetic vessel.

\subsection{Autoencoder Architecture}\label{Autoencoder_Architecture}

This section provides the details of the architecture of Autoencoder used for feature extraction in this study, as shown in Table \ref{tab:DAA}.

\begin{table}[htbp]
\centering
\small
\caption{Detailed architecture of autoencoder}
\label{tab:DAA}
\begin{tabular}{c c c c c}
\hline
L\#      & Type  & Kernel            & Stride           &Activation Function\\ \hline
1       & Input         &                  &                      \\ 
2       & Conv1D        & $32\times3\times1$ & $1\times1$     & ReLU       \\
3       & Conv1D        & $32\times3\times1$ & $1\times1$     & ReLU      \\
4       & MaxPool1d     & $2\times1$         & $2\times1$            \\
5       & Conv1D        & $128\times3\times1$ & $1\times1$     & ReLU       \\
6       & Conv1D        & $128\times3\times1$ & $1\times1$     & ReLU      \\
7       & MaxPool1d     & $2\times1$         & $2\times1$            \\
8       & Conv1D        & $256\times3\times1$  & $1\times1$     & ReLU       \\
9       & Conv1D        & $256\times3\times1$ & $1\times1$     & ReLU      \\
10       & MaxPool1d     & $2\times1$         & $2\times1$            \\
11       & Conv1D        & $32\times3\times1$ & $1\times1$     & ReLU       \\
12       & Conv1D        & $32\times3\times1$ & $1\times1$     & ReLU      \\
13       & MaxPool1d     & $2\times1$         & $2\times1$            \\
14       & Conv1D        & $6\times3\times1$ & $1\times1$     & ReLU       \\
15       & Conv1D        & $6\times3\times1$ & $1\times1$     & ReLU      \\
16       & ConvTranspose1d     & $2\times1$   & $2\times1$            \\ 
17       & Conv1D        & $32\times3\times1$  & $1\times1$     & ReLU       \\
18       & Conv1D        & $32\times3\times1$ & $1\times1$     & ReLU      \\
19       & ConvTranspose1d     & $2\times1$   & $2\times1$            \\ 
20       & Conv1D        & $256\times3\times1$ & $1\times1$     & ReLU       \\
21       & Conv1D        & $256\times3\times1$ & $1\times1$     & ReLU      \\
22       & ConvTranspose1d     & $2\times1$   & $2\times1$            \\ 
23       & Conv1D        & $128\times3\times1$ & $1\times1$     & ReLU       \\
24       & Conv1D        & $128\times3\times1$ & $1\times1$     & ReLU      \\
25       & ConvTranspose1d     & $2\times1$   & $2\times1$            \\ 
26       & Conv1D        & $32\times3\times1$ & $1\times1$     & ReLU       \\
27       & Conv1D        & $32\times3\times1$ & $1\times1$     & ReLU      \\
28       & Conv1D        & $4\times3\times1$ & $1\times1$     & ReLU      \\

\hline
\end{tabular}
\end{table}

\subsection{Feature Fusion U-Net Architecture}\label{FFU_Architecture}

In this section, the architecture details of the FFU-Net are presented in Table \ref{tab:DFFU}, which is used to bridge the feasibility gap between high- and low-fidelity data.

\begin{table}[htbp]
\centering
\small
\caption{Detailed architecture of feature fusion U-Net}
\label{tab:DFFU}
\begin{tabular}{c c c c c c}
\hline
$L\#$      & Type  & Kernel            & Stride           &Activation Function & Comments\\ \hline
1       & Input         &                  &          &           & \\ 
2       & Conv1D        & $32\times3\times1$ & $1\times1$     & ReLU    &   \\
3       & Conv1D        & $32\times3\times1$ & $1\times1$     & ReLU     & \\
4       & MaxPool1d     & $2\times1$         & $2\times1$            &\\
5       & Conv1D        & $64\times3\times1$ & $1\times1$     & ReLU      & \\
6       & Conv1D        & $64\times3\times1$ & $1\times1$     & ReLU    &  \\
7       & MaxPool1d     & $2\times1$         & $2\times1$           & \\
8       & Conv1D        & $128\times3\times1$  & $1\times1$     & ReLU     &  \\
9       & Conv1D        & $128\times3\times1$ & $1\times1$     & ReLU     & \\
10       & MaxPool1d     & $2\times1$         & $2\times1$           & \\
11       & Conv1D        & $256\times3\times1$ & $1\times1$     & ReLU     &  \\
12       & Conv1D        & $256\times3\times1$ & $1\times1$     & ReLU    &  \\
13       & \begin{tabular}{@{}c@{}}Flatten \& \\ Feature Fusion\end{tabular}    &     &      &   &      2096 neurons              \\
14       & Dense Layer   &     &       &  &      2048 neurons              \\
15       & Dense Layer   &     &    &     &      2048 neurons              \\
16       & Reshape       &     &    &  & \begin{tabular}{@{}c@{}}Feature Maps \\ ($256\times1\times16$)\end{tabular}
       \\
17       & ConvTranspose1d     & $2\times1$   & $2\times1$    &      &  \\ 
18       & Conv1D        & $128\times3\times1$  & $1\times1$     & ReLU    &  Skip Connection $L\#9$ \\
19       & Conv1D        & $128\times3\times1$ & $1\times1$     & ReLU    &  \\
20       & ConvTranspose1d     & $2\times1$   & $2\times1$        &    \\ 
21       & Conv1D        & $64\times3\times1$ & $1\times1$     & ReLU  &     Skip Connection $L\#6$\\
22       & Conv1D        & $64\times3\times1$ & $1\times1$     & ReLU  &    \\
23       & ConvTranspose1d     & $2\times1$   & $2\times1$        &    \\ 
24       & Conv1D        & $32\times3\times1$ & $1\times1$     & ReLU  &     Skip Connection $L\#3$\\
25       & Conv1D        & $32\times3\times1$ & $1\times1$     & ReLU    &  \\
26       & Conv1D        & $2\times3\times1$ & $1\times1$     & ReLU   &   \\

\hline
\end{tabular}
\end{table}

\end{appendices}

\section*{Acknowledgments}
This work is supported by the National Science Foundation (NSF) Grant \# 2151555.

\section*{Declarations}

\textbf{Conflict of interest}: The authors declare that they have no conflict of interest.

\textbf{Replication of results}: The source code of this study is implemented using python. All codes are available from the author upon reasonable request.

\bibliography{sample}
\end{document}